\documentstyle[aps,prl,epsfig,multicol]{revtex}

\begin{document}

\title{Slave-Boson Three-Band Model with O-O Hopping for High-T$_c$ Superconductors }

\author{Ivana Mrkonji{\' c} and Slaven Bari{\v s}i{\' c}}
\address{Department of Physics, University of Zagreb, Bijeni{\v c}ka 32, POB 331, 10002 Zagreb,
Croatia}
\date{3 June 2002}
\maketitle

\begin{abstract}
Slave boson mean-field approximation is carried out analytically
for weakly doped Cu$O_2$ conduction planes, characterized by Cu-O
charge transfer energy $\Delta_{pd}$, Cu-O hopping $t_0$, O-O
hopping $t'$ and repulsion $U_d$ between holes on Cu site taken as
infinite. At zero doping $\delta$, finite negative
t',$|t'|<t_0/2$, expands the range of stability of the covalent,
conducting state on the expense of the insulating state which,
however, remains stable at larger $\Delta_{pd}$. For sufficiently
large $\Delta_{pd}$ the renormalized charge transfer energy
saturates at $4|t'|$ instead of decreasing to zero, as at $t'=0$
case. In contrast to $t'$, finite $\delta$ suppresses the
insulating state nearly symmetrically with respect to the sign of
$\delta$. The regime with charge transfer energy renormalized
close to $4|t'|$ fits remarkably well the ARPES spectra of Bi2212
and LSCO, and, in the latter case, explains the observed strong
doping dependence of the Cu-O hopping.
\end{abstract}


\begin{multicols}{2}
 ARPES measurements of the electron spectra in the high-T$_c$
superconductors \cite{Ino,Norman,MCA,Shabel1} have shown that the
characteristic band energy scales fall into the range between 0.1
and 1 eV. On the other hand, the high-energy spectroscopies
indicate considerably larger characteristic energies, of the order
of several eV \cite{Plakida}. This implies that the
renormalization of the electron spectrum in low energy range
measured by ARPES is strong. Recent results on
La$_{2-x}$Sr$_x$CuO$_4$ (LSCO)\cite{Ino} have shown in addition
that the shape of the Fermi surface varies strongly with doping,
indicating again strong renormalization of band parameters.
Further argument in favor of the strong renormalization comes from
the approximate electron-hole symmetry of the phase diagram of the
high-T$_c$ superconductors \cite{Plakida}. If the couplings were
weak, band renormalization small, the approximate electron-hole
symmetry would occur with respect to the (optimal) doping
$\delta_c$, necessary to bring the Fermi level to the logarithmic
van Hove singularity. This would be accompanied by the activation
of the Umklapp scattering, i.e. tendency towards Mott
localization. Thus far such tendency has not been observed and the
approximate symmetry occurs with respect to $\delta=0$ rather than
with respect to $\delta_c$. This rules out weak coupling and
raises the question whether intermediate \cite{Friedel} to strong
\cite{Kotliar}coupling can account for such behavior.

    It is usually assumed that the largest coupling is the local
hole-hole interaction $U_d$ on the Cu-site of the conducting
CuO$_2$ planes. The interaction problem is often treated within
the Emery model \cite{Emery} which describes the three-site
structure of the CuO$_2$ plane by copper and oxygen site energies
$\varepsilon_d$ and $\varepsilon_p$ respectively, the Cu-O hopping
$t_0$ and the O-O hopping $t'_0$, the single-particle parameters
in addition to U$_d$. The strong renormalization of the band
structure characterized by $\varepsilon_d$, $\varepsilon_p$, $t_0$
and $t'_0$ is thus expected only when the number of holes on the
Cu-site is close to unity, because only then large U$_d$ is
effective. In the opposite case, when the holes tend to reside on
oxygen sites, the renormalization should decrease. Bearing this in
mind, the limit of large U$_d$ is discussed here, taking $U_d$
larger than $\Delta_{pd}$= $\varepsilon_p - \varepsilon_d$, t$_0$,
t'$_0$. The limit of infinite U$_d$ is, as usual, treated within
mean-field slave boson approximation \cite{Kotliar,Grilli},
neglecting the AF and SC effects. The latter occur \cite{Plakida}
on the energy scales of the 0.01 eV, an order of magnitude below
the band energy scales of interest here.

Previously, similar slave boson calculations were carried out
analytically for the $t_0$, $t'_0$ 1d (CuO) analog of the 2d
(Cu$O_2$) case \cite{Grilli}, and, for the latter, numerically for
some particular choices of the bare parameters \cite
{Grilli,Tutis,Golosov}. The 2d three-band structure is
considerably richer than its 1d two-band counterpart, therefore,
similar holds for the corresponding mean-field slave boson theory.
In particular, the 2d case is sensitive to the sign of $t'_0$. Two
bands may anticross, as in  the 1d case, but in 2d case,
band-touching is also possible \cite{Golosov,im}. Above all, in
contrast to 1d model, the parameter $t'_0$ removes in 2d case
Fermi energy of the half-filled band from the van Hove
singularity, i.e. introduces a finite doping $\delta_c$ required
to bring it back \cite{im}. The question whether the electron-hole
symmetry occurs for $\delta=0$ or $\delta_c$ should therefore be
discussed at finite t'$_0$.

The slave boson procedure \cite{Kotliar} amounts to searching the
ground state band energy $E_0$ at a given number of holes $1+
\delta$, subjected to the restriction that the number of holes
$n_d$ on the Cu-site is smaller than 1, as required by large
$U_d$. In the mean-field approximation $E_0$ is the energy of the
free holes with the renormalized band parameters $\Delta, t, t'$,
chosen to fulfill the requirement $\langle n_d \rangle <1$, which
becomes \cite{Kotliar}
\begin{equation}
\label{e1}
 \langle n_d \rangle + t^2/t_0^2=1, \:\:\langle n_d \rangle =-\frac{\partial E_0}{\partial
\Delta}.
 \end{equation}

The minimization of $E_0$ with respect to the remaining parameter
t, gives the second slave boson equation
\begin{equation}
\label{e2}
 \langle n_B \rangle =2 \frac{t(\Delta_{pd}- \Delta)}{t{_0^2}},\:\:
 \langle n_B \rangle =\frac{\partial E_0}{\partial
t},
\end{equation}
denoting renormalized bond charge on the Cu-O bond.

The solution of the coupled integro-differential equations
(\ref{e1}) and (\ref{e2}) gives $\Delta$ and t as functions of
$\Delta_{pd}$, $t_0$, t' and $\delta$ (remembering that
$t'_0=t'$). The present slave boson calculation is carried out for
arbitrary $\Delta_{pd}$ and $t_0$, assuming $2|t'|<t_0$ and
$|\delta|<<1$. The assumption $2|t'|<t_0$ is in the spirit of the
Emery model \cite{Emery}, but the opposite limit will nevertheless
be briefly commented upon. On the other hand, the results obtained
here for $|\delta|<<1$ do not cover all the physically achieved
doping, but rather indicate the tendencies associated with finite
$\delta$.

It is possible to transform \cite{im1} the integro-differential
equations (\ref{e1}) and (\ref{e2}) to the coupled algebraic
equations for all interesting regimes of parameters by calculating
$E_0(\Delta,t,t', \delta)$ analytically and then $\langle n_d
\rangle$ and $\langle n_B \rangle$ from Eqs. (\ref{e1}) and
(\ref{e2}). Leaving the details of the calculation for the
extended publication \cite{im1}, the final results will be
discussed here.

These results are best understood starting from the $t'=0$
situation. $\varepsilon_p$, which does not renormalize, is chosen
for the energy origin, $\varepsilon_p=0$. Only one band,
$\varepsilon_L^0(\bf k)$, out of three is occupied for $\delta
\leq 1$,
\begin{eqnarray}\label{e3}
  \varepsilon_L^0({\bf
  k})=-\frac{1}{2}(\Delta_0+\sqrt{\Delta_0^2+16t^2f_1}),\nonumber \\
f_1 = \sin^2\frac{k_{x}}{2}+\sin^2\frac{k_{y}}{2},
\end{eqnarray}
setting $\Delta(t'=0)=\Delta_0).$ Using $\varepsilon_L^0({\bf k})$
to calculate $E_0$, leads to the solution of the t'=0 slave boson
equations (\ref{e1}) and (\ref{e2}) for arbitrary $\Delta_{pd}$
and $t_0$ (taken positive without loss of generality). The
behavior of t for $\delta=0$ is fairly simple: $t/t_0=0$, starting
from large $\Delta_{pd}$ to $\Delta{_{pd}^{cr}}=4.7t_0$, where the
angular point in $t/t_0$ occurs. For $\Delta_{pd}$ close below
$\Delta{_{pd}^{cr}}$, t is given by a Landau-like expression
$t\sqrt{\Delta{_{pd}^{cr}}}=1.15t_0\sqrt{\Delta{_{pd}^{cr}}-\Delta_0}$,
which describes Brinkman-Rice (BR) phase transition from the
insulating $t=0$ to the conducting phase. The overall behavior can
be obtained by expanding Eq.(\ref{e3}) in terms of
$t^2/\Delta{_0^2}$ including the terms of the order of $t^4$ close
below $\Delta{_{pd}^{cr}}$ \cite{Kotliar}.

The corresponding behavior of $\Delta_0(\Delta_{pd})$ is more
important for further discussion. $\Delta_0$ has a single, maximum
(angular at $\delta=0$) at $\Delta_{pd}=\Delta{_{pd}^{cr}}$, where
it reaches the value $\Delta_0=\Delta{_{pd}^{cr}}/2$. For
$\Delta_{pd}$ well below $\Delta{_{pd}^{cr}}$ the regime
$t>|\Delta_0|$ is reached. $\langle n _d \rangle$ decreases,
according to Eq.(\ref{e1}) and so do both renormalisation of $t_0$
to t and of $\Delta_{pd}$ to $\Delta_0$. Before approaching
asymptotically $\Delta_{pd}$ for the large negative
$\Delta_{pd}/t_0$, $\Delta_0$ crosses zero \cite{Tutis} at
$\Delta_{pd}/t_0 \approx 1.5$: standard covalent situation
$\Delta_{pd} \approx t_0$ renormalizes to the extreme $\Delta_0
\approx 0$ covalent limit. For $\Delta > \Delta{_{pd}^{cr}}$,
$\Delta(\Delta_{pd})$ decreases fast towards the asymptotic regime
valid for large $\Delta_{pd}$, when $\Delta=4.7t_0^2/\Delta_{pd}$
tends to zero (similarly to the site energy of the t-J model
\cite{Rice1}).

Turning next to finite t', in the limit where $2|t'|<t_0$, it is
to be noted that under such condition (more precisely for
$4|t'|<\Delta{_{pd}^{cr}}/2=2.4t_0$), $4|t'|$ intersects
$\Delta_0(\Delta_{pd})$ at $\Delta'_{pd}>\Delta{_{pd}^{cr}}$. When
$|t'|\ll t_0$, $\Delta'_{pd}=4.7t_0^2/|t'|$. For
$\Delta_{pd}<\Delta{_{pd}^{'}}$, $4|t'|$ is smaller than
$\Delta_0$ and/or t obtained from t'=0 solution, and a
perturbative treatment of $t'$ is possible. For
$\Delta_{pd}>\Delta_{pd}'$, $4|t'|>\Delta_0$ and full treatment of
t' is required.

    The perturbative corrections due to t' in the regime
$t>\Delta_0$ at $\Delta_{pd}$ well below 4.7$t_0$, are not of
particular interest because only of the quantitative nature. The
attention will be therefore focused here on the small t regime at
$\Delta_{pd}$ close below or above 4.7$t_0$. For this purpose a
perturbative calculation can be carried out in terms of small t,
using (with $\varepsilon_p=0$)
\begin{eqnarray}
\label{e4} \varepsilon_L({\bf k})= -\Delta - 4t^2 \frac{\Delta f_1
- 8t'f_2 }{\Delta^2 - 16t'^2 f_2} + O(t^4), \nonumber \\
f_1 = \sin^2\frac{k_{x}}{2} \sin^2\frac{k_{y}}{2}.
\end{eqnarray}
This expansion, valid at $\delta \approx 0$ for $0<t<\Delta -
4|t'|$ and arbitrary sign of $t'$, reduces at t'=0 to the spectrum
used in obtaining the BR transition, i.e it is appropriate for
following the evolution with t' of the small t slave boson
solution.

In the first step Eq.(\ref{e4}) is used in the range
$4.7t_0<\Delta_{pd}<\Delta'_{pd}$, where it can be expanded
further in terms of $0<|t'|/\Delta<1$. The following results are
obtained retaining the terms linear in t' and quadratic and
quartric in t.

    The corresponding analytical calculation of the position of the Fermi level $\varepsilon_F$ has been
carried out earlier \cite{im}, by systematical linearization with
respect to t'. The shift of the $\varepsilon_F$ from the van Hove
singularity at $\Delta_0
(\varepsilon_{vH}-\varepsilon_F)=2t_0^2x_F$ turned out to be
\begin{equation}
\label{e5}
  x_F\ln\frac{1}{x_F}=\frac{8\pi^2(\delta_c - \delta)}{1+8\pi^2\delta_c}
\end{equation}
where $\delta_c=-32t'/\pi^2\Delta_0$ stands for the critical
doping required to bring the Fermi level back to the van Hove
singularity. Experimentally, this doping is positive i.e. ${\rm
sign}t'= - {\rm sign}\Delta$. The linearization procedure can be
generalized to the calculation of $E_0(\Delta,t, t',\delta) $, and
then, straightforwardly, to the solution of the slave boson
equations (\ref{e1}) and (\ref{e2}). Here, for brevity, small
terms in $E_0$, linear in $x_F$, discussed in Ref. \cite{im1},will
be omitted in the description of this solution.

The BR transition is shifted to the new position
$\Delta_{pd}^{cr}=4.7t_0+3.5 \delta_c$, with
$\delta_c=13.6t'/t_0$, but it is not suppressed by t'. This
important result can be traced back to the outstanding feature of
the expansion of Eq.(\ref{e4}) in terms of $t'/\Delta$, generating
the term $t^2t'/\Delta^2$. The above term vanishes with t, making
the theory all along convergent with respect to $t'/\Delta$, as
exemplified by Eq.(\ref{e5}), where $x_F$ is finite and small
independently on t.

$\Delta_{cr}$, the value of $\Delta$ in the angular point, is
equal to the $\Delta{_{pd}^{cr}}/2$, i.e. for $\delta_c>0$ it is
slightly increased with respect to the $t'=0$ value, and $\Delta$
remains above $\Delta_0=\Delta(t'=0)$ in the whole range of
convergence of this
approach,$\Delta{_{pd}^{cr}}<\Delta_{pd}<\Delta'_{pd}$.

These results can be extended to finite $\delta>0$. In contrast to
$t'$, i.e. to $\delta_c$, finite $\delta$ suppresses the BR
transition, turning it into crossover, in spite of Eq.(\ref{e5}),
which suggests analogous roles of $\delta_c$ and $\delta$. In the
vicinity of $\Delta{_{pd}^{cr}}$, the angular behavior of $\Delta$
is essentially modified by finite $\delta$. For example, the value
of $\Delta$ at $\Delta_{pd}=\Delta_{pd}^{cr}=4.7t_0+3.5\delta_c$
is

\begin{equation}
\label{e6}
 \frac{\Delta{_{pd}^{cr}}}{2}-\Delta =0.8(1+0.2\delta_c)\delta^{1/3}
\end{equation}
and correspondingly
\begin{equation}
\label{e7}
  t^2/t_0^2=0.63(1-0.2\delta_c)\delta^{2/3},
\end{equation}
up to the leading orders in $\delta$ and $\delta_c$.

Finally we turn to the range $\Delta_{pd}>\Delta'_{pd}$, where the
stability of the $t=0$ solution is to be discussed using
Eq.(\ref{e4}) non-expanded in terms of t', but dropping out
$O(t^4)$ term. The main contribution to the cohesive energy $E_0$
is weakly dependent on $\varepsilon_F$ (i.e. $\delta_c$) and
strongly dependent on lower cut-off, which is
$\widetilde\Delta=\Delta-4|t'|$ according to Eq.(\ref{e4}), i.e.
$\delta_c$. (Note in this respect that the $\varepsilon_F$
dependency of E$_0$ was not very important even in the vicinity of
the BR transition, Eqs.(\ref{e6})and (\ref{e7}).) Expanding then
Eq.(\ref{e4}) around ${\bf k}=(\pi, \pi)$ point, one finds
immediately that $E_0$ is logarithmic in $t^2/t'\widetilde\Delta$,
which, through slave boson equations, leads for $\Delta_{pd} \gg
\Delta'_{pd}$ at $\delta=0$ to
\begin{equation}
\label{e8}
  \widetilde\Delta=\Delta-4|t'| \approx
  2|t'|e^{-\frac{|t'|\Delta_{pd}}{4c t{_0^2}}},
\end{equation}
where c is a numerical factor. Eq.(\ref{e8}) shows that the
asymptotic behavior of $\Delta$ which at $t'=0$ vanishes as
$\Delta_0 \sim t_0^2/\Delta_{pd}$, undergoes an essential
modification to the exponential behavior, with the saturation at
$\Delta=|4t'|$. Although the derivation of the result (\ref{e8})
does not suggest that this result is independent of the dimension,
the similar behavior was obtained \cite{Grilli} at $\delta=0$ in
1d. Moreover, it is easy to see that finite $\delta>0$ leads to $t
\sim \sqrt{\delta}$ as before. However, the convergence of this
result is reduced to $t<\widetilde\Delta$, which is itself
(exponentially) small according to Eq.(\ref{e8}). The regime $t
\approx \widetilde\Delta$ is quickly reached with doping and it is
clear that,from the slave boson point of view, the limit $t \geq
\widetilde\Delta$ is physically achievable.

The corresponding fits of the ARPES data for the Fermi surfaces of
LSCO at different doping $\delta$ are shown in Fig.1. These fits
also require ${\rm sign}\Delta =- {\rm sign} t'$ and determine two
parameters e.g. $\Delta/t'$ and $t/t'$ as functions of $\delta$.
While $\Delta \approx 4|t'|$ is almost independent on doping, t
increases quickly with doping. For the underdoped regime (small
doping) the behavior of t can be approximated by $ \sqrt{\delta}$,
corresponding to the heavy-fermion limit, as suggested by
here-presented mean-field slave boson theory. However, for larger
doping, appreciable values of t are reached, leading the system to
the essentially covalent behavior. Finally, for overdoped regime,
t and $\Delta$ seem to saturate, requiring further theoretical
investigation, in particular because the 1d theory \cite{Grilli}
suggests otherwise. Here, the calculus becomes quite complicated
since the perturbative treatment of $t$ is no longer possible, and
the exact hole spectra are required. Fig.\ref{fig2}a shows the
conducting band fit of Bi$_2$Sr$_2$CaCu$_2$O$_8$ (Bi2212), which
fixes the absolute value of t in addition to the relative values
obtained from the Fermi surface fits. t turns out to be of the
order of 0.1eV, almost an order of magnitude less than t$_0$,
again in agreement with the present slave boson analysis.

\vspace*{0.1cm}
\begin{figure}
\begin{center}
\epsfxsize=3in \epsfbox{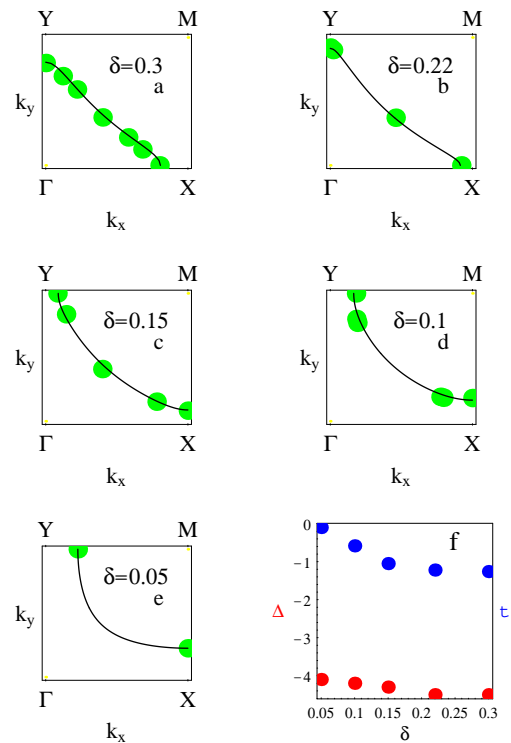}
\end{center}
\caption{(a-e) Experimentally measured Fermi surfaces for
LSCO\protect\cite{Ino} (green dots) and the three band model fits
(solid). (f) fitting parameters in units of $t'$ with
$\widetilde\Delta$ small}
 \label{fig1}
\end{figure}
\vspace*{0.1cm}

Equally good fits (c.f. Fig.\ref{fig2}b) were however obtained
earlier \cite{im} assuming $|\Delta|<4|t'|$ rather than
$|\widetilde\Delta|<4|t'|$. Since it was shown here that the
latter is the solution of the $\delta \approx 0$ slave boson
theory when $2|t'|<t_0$, the former can only occur at $\delta
\approx 0$, if $2|t'|>t_0$. While the behavior of $\widetilde
\Delta$ given by Eq.(\ref{e8}) can be visualized, assuming for the
moment $t=0$, as the avoided crossing of the dispersionless (t=0)
copper level and the pure oxygen band with the energy $-4|t'|$ at
${\bf k}=(\pi, \pi)$ point, the limit $|\Delta|<4|t'|$ means that
the Cu level has entered the oxygen band. The holes on Cu site are
then necessarily transferred to the oxygen sites, i.e. $\langle
n_d \rangle < 1$. For $n=1+ \delta$, with $\delta \approx 0$,
Eq.(\ref{e1}) then requires finite t, i.e the solution t=0 is
unstable for $2|t'|>t_0$. (For finite t, copper band undergoes
anticrossing (sign t' = - sign $\Delta$), rather than touching
(sign t' = sign $\Delta$) with the oxygen band.\cite{Golosov,im})
With $\langle n_d \rangle<1$, the renormalization becomes small,
analogously to the case of small and negative $\Delta_{pd}$ at
t'=0. $t \leq t_0$ and $\Delta \leq \Delta_{pd}$ are not expected
to depend much on doping $\delta$. The appreciable experimental
dependence of t on small $\delta$ \cite{Ino} is therefore an
indication in favor of the strongly renormalized $2|t'|<t_0$
solution.

 \vspace*{0.1cm}
\begin{figure}
\begin{center}
\epsfxsize=3in \epsfbox{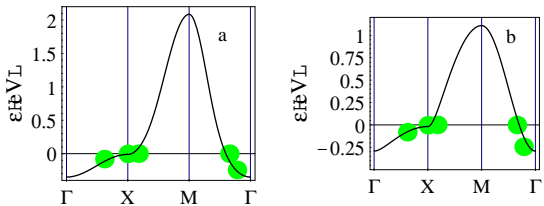}
\end{center}
\caption{Band-fits for the conducting band of Bi2212 of
\protect\cite{Norman}, (a) $\widetilde \Delta$ small, (b) $\Delta$
small}\label{fig2}
\end{figure}
\vspace*{0.1cm}

Let us finally point out that $2|t'|<t_0$ solution, strongly
dependent on doping, is approximately symmetric with respect to
the electron ($\delta < 0$) or hole doping ($\delta> 0$). As the
latter was discussed above, leading in particular to $t/t_0 \sim
\sqrt{\delta}$ for sufficiently large $\Delta_{pd}>4.7t_0$, the
case $\delta<0$ will be briefly mentioned here. Since $\langle
n_d\rangle \leq n = 1+ \delta$, Eq.(\ref{e1}) means that $t\geq
t_0\sqrt{-\delta}$, i.e., roughly speaking, $t$ behaves as
$\sqrt{|\delta|}$, irrespectively of the sign of $\delta$. $t=0$
line of the slave boson theory, obtained for the half-filled
$\delta=0$ Cu-based band, is the singular line of the phase
diagram, not only at t'=0, but also for $t' \neq 0$, as shown
here. In the properties depending on the details of the band
structure close to the Fermi surface, the simplest of which is
perhaps the Hall constant, the symmetry is broken when a finite
t'is taken into account. The corresponding asymmetry is the
largest in the small $t'/\Delta$ limit of Eq.(\ref{e5}), when
$\delta>0$ means doping towards and $\delta<0$ away from the vH
singularity, whereas in the opposite limit of Eq.(\ref{e8}), the
Fermi level at $\delta=0$ is so far from the vH singularity, that
the variation of the topology of the Fermi surface with doping
becomes small. Overall, the asymmetry is much smaller than in the
weak coupling limit.

In conclusion, the present work shows that the band structure of
the high-T$_c$ superconductors can be understood in terms of p-d
bands in the CuO$_2$ plane, renormalized by strong interaction on
the Cu- site. According to the low energy ARPES data, the bare
band parameters, measured by high energy spectroscopy, are
renormalized by an order of magnitude. More profound understanding
of the energy scales, of the order of 0.1 eV, i.e. the
understanding of the approximate ground state on these energy
scales, represents an appropriate prerequisite for consideration
of the symmetry breaking, associated with AF and SC, on the energy
scales of 0.01 eV.

\begin{acknowledgments}
Thanks to J. Friedel, L. P. Gork'ov and Ph. Nozi{\` e}r{\` e}s for
useful discussions. This work was supported by Croatian Ministry
of Science under the project 119-204.
\end{acknowledgments}

\end{multicols}

\end{document}